


\magnification=\magstephalf
\hoffset=0.5in
\voffset=0.2in
\input amstex
\documentstyle{amsppt}
\pagewidth{29pc}
\NoBlackBoxes

\leftheadtext{R. I. MCLACHLAN AND H. SEGUR}
\rightheadtext{A NOTE ON THE MOTION OF SURFACES}

\topmatter
\title A note on the motion of surfaces
\endtitle
\author Robert I. McLachlan and Harvey Segur\endauthor
\date June 22, 1993 \enddate
\address Program in Applied Mathematics, University of Colorado at Boulder,
Boulder, CO 80309-0526
\endaddress
\email rxm\@boulder.colorado.edu, segur\@boulder.colorado.edu\endemail

\font\bb=msym10
\def\RR {\hbox{\bb R}}
\font\mib=cmmib10
\def\bld#1{\hbox{\textfont1=\mib$#1$}}
\def\ii{{i'}}
\def\r{\bold r}
\def\x{\bold x}
\def\t{\bld{\tau}\!}
\def\n{{\bold n}}
\def\bt{{\bold t}}
\def\b{{\bold b}}
\def\c{\boldkey ,}
\def\s{\boldkey ;}
\loadbold

\abstract
We study the motion of surfaces in an intrinsic formulation in which the
surface is described by its metric and curvature tensors. The evolution
equations  for the six quantities contained in these tensors are reduced in
number in two cases:
(i) for arbitrary surfaces, we use principal coordinates to obtain two
equations for the two principal curvatures, highlighting the similarity
with the equations of motion of a plane curve;
and (ii) for surfaces with
spatially constant negative curvature, we use parameterization by
Tchebyshev nets to reduce to a single evolution equation. We also
obtain necessary and sufficient conditions for a surface to maintain
spatially constant negative curvature as it moves. One choice for
the surface's normal motion leads to the modified-Korteweg de Vries equation,
the appearance of which is explained by connections to the AKNS
hierarchy and the motion of space curves.

\endabstract

\endtopmatter

\document
\def\headfont@{\bf}

\head 1. Introduction and equations of motion
\endhead

Studies of the motion of surfaces and interfaces occur in several
scientific disciplines, including geometry [8,10,17], water waves [20], crystal
growth [11,13], combustion [4], and gas dynamics [18].  Pelc\'e [16]
collected and
reprinted an interesting collection of original articles in some of these
areas.  To simplify the analysis, these studies often consider only
two-dimensional motion, thus reducing the problem to the motion of a curve
[6], but this is only a special case of the usual situation.

To study the evolution of two-dimensional
surfaces arbitrarily embedded in three dimensions, a natural approach
is to describe the surface purely intrinsically, that is, in terms of
the surface's metric and curvature tensors.  In these variables, the
equations of motion for the surface take the form of six coupled nonlinear
evolution equations for the six independent components of the two tensors.
These equations were first written down by Brower {\it et al.} [3], and
later by Nakayama \& Wadati [15].

Unfortunately, these equations are highly redundant.  To see this, observe
that if a surface that can be represented at a fixed time in the form
$z=z(x,y)$, then its motion can be described by a single evolution equation,
for $\partial z/\partial t$.
The purpose of this paper is to simplify the evolution
equations obtained in [3] and [15], by reducing the number of unknowns.
We obtain two simplifications, by choosing two different parameterizations
of the surface.

The first parameterization, in \S2, is general:  we parameterize a surface
by its principal lines
({\it i.e.}, lines tangent to the directions of principal curvature). This
reduces the
system to four coupled equations, and, with a particular choice of
parameterization
along the principal lines, to two equations for the two curvatures.  If the
surface happens to be a cylinder above a curve in a plane, and if the
motion is two-dimensional, then one curvature vanishes and the
equations reduce to the single equation obtained in [3,14] for the evolution of
the curvature of a plane curve.

The second parameterization, in \S3, is more specialized: surfaces of
constant negative
curvature have an elegant parameterization by Tchebyshev nets.
For surfaces that maintain constant negative curvature as they move, their
motion is specified by a single evolution equation for the angle between
the coordinate lines.  As an example, we show that one choice for the
normal velocity of the surface
leads to the modified Korteweg-de Vries equation for the surface's evolution.

For either set of coordinates discussed here, the key idea is to force the
coordinate lines to maintain
their defining property as the surface evolves, by adding an appropriate
tangential velocity to the evolution.

The first step is to derive the six coupled equations for the motion of a
surface moving in $\RR^3$. Local coordinates on the surface
are $(x_1,x_2)$ = $\x$, time is $t$, and the position (in $\RR^3$) of a
point on the surface
is $\r(\x,t)$.  Later $\dot\r$ ($=\partial\r/\partial t)$ will be specified, so
one
can think of $\x$ as a Lagrangian particle label on the surface.
Define two tangent vectors and one normal vector:
$$ \t_\mu=\r_{\c\mu}\qquad \n={\t_1\times\t_2\over|\t_1\times\t_2|},$$
where greek indices range over $\mu=1,2$ with
$_{\c\mu}={\partial\over\partial x_\mu}$. (Both the notation used
here and the preliminary development follow Spivak [19, Chapter 2].)
Note that $\n$ is a unit
vector but the $\t_\mu$ are not necessarily unit vectors.
We have the metric $g_{\mu\nu}$ and curvature $h_{\mu\nu}$ tensors (which
define
the first and second fundamental forms $g_{\mu\nu}dx_\mu dx_\nu$ and
$h_{\mu\nu}dx_\mu dx_\nu$, respectively):
$$ g_{\mu\nu}=\t_\mu\cdot\t_\nu,\qquad
   h_{\mu\nu}=\n\cdot\t_{\mu\c\nu}=\n\cdot\r_{\c\mu\nu}\eqno(1.1)$$
As one moves along the surface (at a fixed time), the tangent and normal
vectors change
according to the Gauss-Weingarten equations,
$$ \eqalign{\t_{\mu\c\nu}&=\t_\lambda\Gamma^\lambda{}_{\!\mu\nu}+\n
h_{\mu\nu} \cr
            \n_{\c\nu}&=-\t_\lambda g^{\lambda\mu}h_{\mu\nu} \cr}\eqno(1.2)$$
where $\Gamma^\lambda{}_{\!\mu\nu}$ are the Christoffel symbols of the
second kind
defined by the metric $g_{\mu\nu}$, and $g^{\mu\nu}=(g_{\mu\nu})^{-1}$
as usual. Repeated indices are summed on unless otherwise noted.

The six quantities in (1.1) are not all independent.  They are related by three
consistency conditions for the PDE's (1.2)---%
the Gauss-Codazzi equations:
$$\eqalignno{ R_{1212}&=\hbox{det} h&(1.3a)\cr
	    h_{\nu\lambda\s\mu}&=h_{\mu\lambda\s\nu}&(1.3b)}$$
where $_{\s\mu}$ denotes covariant derivative and
$R_{1212}$ is the nontrivial element of the Riemann tensor. Note that
(1.3b) gives independent information only for
$(\nu,\lambda,\mu)=(1,1,2)$ or $(2,2,1)$. Of the three remaining
degrees of freedom, two are due to the arbitrary parameterization, leaving
one---which we expect, since the surface could be written, {\it e.g.},
$\r=(\x,z(\x),t)$.

Therefore one might ask: What is to be gained by going to an ``intrinsic''
({\it i.e.} $g_{\mu\nu}$, $h_{\mu\nu}$) representation of the surface when eqs.
(1.3) cannot
usually be solved explicitly? Any surface motion can after all
be formulated locally as $\dot z=f(\x,t)$.
Such an approach fails if the surface becomes vertical---%
an apparently artificial constraint---but the intrinsic formulation can
also suffer from artificial coordinate singularities.  One advantage is that
in some applications,
the motion of the surface normal to itself is given naturally in terms of the
local curvatures [3,17]
which are the eigenvalues of $g^{\mu\lambda}h_{\lambda\nu}$;
secondly, such an approach has been successful in the study
of the motion of curves, as we now discuss.

Hasimoto [9] showed that in the self-induction approximation,
the motion of a 3D vortex filament
(which moves in the direction of its local binormal vector with speed equal
to its local
curvature) can be reduced to  the nonlinear Schr\"odinger equation. Lamb [12]
extended this to more general motions of curves, and found motions obeying the
sine-Gordon and modified Korteweg-de Vries equations.  All three of these
equations are completely integrable [1], and Goldstein and
Petrich [7] showed how to specify infinitely many two-dimensional motions of
curves in terms of integrable evolution equations.  The occurrence of such
integrable equations
was shown in [14] to be due to the fact
that the Serret-Frenet equations,
($\bt=\t\,/|\t\,|$, $s=$arclength)
$$\pmatrix \bt_{\c s}\cr \n_{\c s}\endpmatrix=\pmatrix
0&\kappa\cr-\kappa&0\cr\endpmatrix
\pmatrix \bt\cr\n \endpmatrix\eqno(1.4)$$
which are the analog of (1.2) for plane curves, are equivalent to the AKNS
scattering problem [1] at zero eigenvalue. Thus
evolutions $\pmatrix \dot\bt\cr\dot\n\endpmatrix$ can be specified which
give rise to
integrable equations of the AKNS hierarchy.

The equations for surfaces, (1.2), are more complicated than those for
curves, (1.4), for two main reasons:  (i)  $\bt$ in (1.33) is a unit
vector, whereas the $\t_\mu$ in (1.2) are not; and (ii) the coordinates, $x_i$
, in (1.2) need not measure arclength.  If we relax these two requirements
in (1.4), then they become
$$\pmatrix \t_{\c x}\cr\n_{\c x}\endpmatrix=\pmatrix g_{\c x}/2g&h\cr
-h/g&0\endpmatrix
\pmatrix \t\cr\n \endpmatrix$$
which are equivalent to (1.2) when $\mu=1$ only.

There is another interesting connection between curves and surfaces which
we note briefly. In [14] it is shown that the Serret-Frenet equations
for {\it space} curves also reduce to the AKNS scattering problem at
zero eigenvalue ({\it i.e.,} they take the form (1.4) in appropriate
variables). Considering a moving space curve as sweeping out a surface, we
see that the equations analogous to Serret-Frenet for the surface,
the Gauss-Weingarten equations (1.2), must also contain the AKNS scattering
problem (in appropriate variables). This point of view explains
the occurrence of integrable equations from the AKNS hierarchy in
the equations defining particular surfaces, such as the sine-Gordon
equation for surfaces of constant negative curvature.

Motion of a surface can be prescribed in the form
$$ \dot\r=U\n+V^\mu\t_\mu\eqno(1.5)$$
where $U$ and $V^\mu$ can be specified arbitrarily.
This implies (differentiate $\t_\mu=\r_{\c\mu}$ and $\n\cdot\t_\mu=0$ and use
(1.2)):
$$\eqalign{
\dot{\t}_\mu&=(U_{,\mu}+V^{\nu}h_{\mu\nu})\n +
(-Uh^\lambda{}_{\mu}+V^\lambda{}_{\!\s\mu})\t_\lambda\cr
\dot\n&=-g^{\lambda\mu}(U_{\c\mu}+V^{\nu}h_{\mu\nu})\t_\lambda\cr
}\eqno(1.6)$$
Now intrinsic evolution equations for $\dot{g}_{\mu\nu}$, $\dot{h}_{\mu\nu}$
can be derived in two ways. The first is
to differentiate (1.1) with respect to $t$ and to eliminate $\t_\mu$ and $\n$
using (1.2, 1.3). This was first done by Brower [3] for $V^\mu=0$ and by
Nakayama {\it et al.} [15] for $V^\mu\ne0$. There is also an ``intrinsic''
derivation, namely requiring compatibility of the PDE's (1.2) and (1.6).
This
gives equations which must be solved for $\dot g_{\mu\nu}$ and $\dot
h_{\mu\nu}$,
and the result is (of course) the same:

$$\eqalign{
\dot{g}_{\mu\nu}&=-2h_{\mu\nu}U+V_{\mu\s\nu}+V_{\nu\s\mu}\cr
\dot{h}_{\mu\nu}&=U_{\s\mu\nu}-h_{\mu\lambda}h^\lambda{}_{\nu}U +
h_{\mu\lambda}V^\lambda{}_{\!\s\nu} + h_{\nu\lambda}V^\lambda{}_{\!\s\mu} +
V^\lambda h_{\mu\lambda\s\nu} \cr
}\eqno(1.7)$$

Notice that $V^\mu$ does not affect the actual shape of the surface.  The
easiest way to see this is to note that the evolution in (1.5) is linear
in $U$, $V^\mu$, so we can consider $U=0$;
then the $V^\mu$ velocities just push particles around on a motionless
surface.  Notice also that if a coordinate system with some particular
defining property is chosen initially, that property will be destroyed
under
general motion of the particles on the surface.  Therefore we will use the
tangential velocities $V^\mu$ to ``push'' the coordinates back, so that
they maintain their defining property as the surface moves.  (In the next
sections we show how to do this in
some special cases.)  Nakayama {\it et al.} [15] in addition allowed the
coordinates to evolve on the surface,
but this does not give any extra generality.

\head 2. Principal coordinates
\endhead

The principal curvatures ($\kappa_1$, $\kappa_2$)
and principal directions of a
surface are the eigenvalues and eigenvectors of
$g^{\mu\lambda}h_{\lambda\nu}$. If the coordinate lines are chosen
to be tangent to the principal directions at each point on the surface,
then $g_{\mu\nu}$
and $h_{\mu\nu}$ are diagonal, and we have {\it principal coordinates}.
These can be defined globally on any surface, and are smooth
except where $\kappa_1=\kappa_2$---%
umbilic points. (See [19, p. 288] for more information about
umbilic points; they are generally isolated.)
We define $g_1$, $g_2$, $\kappa_1$, and $\kappa_2$ by
$$ g_{\mu\nu}=\pmatrix e^{g_1}&0\cr0&e^{g_2}\endpmatrix,\qquad
   h_{\mu\nu}=\pmatrix e^{g_1}\kappa_1&0\cr0&e^{g_2}\kappa_2 \endpmatrix. $$
Note that $g_i$ and $\kappa_i$ do not form covariant vectors---they are just
functions appearing in the 2-tensors $g_{\mu\nu}$ and $h_{\mu\nu}$.

In terms of these new variables, the Gauss-Codazzi equations (1.3b) now read
$$ g_{1\c2}={2\kappa_{1\c2}\over \kappa_2-\kappa_1},\qquad
   g_{2\c1}=-{2\kappa_{2\c1}\over \kappa_2-\kappa_1} \eqno(2.1)$$
which we will use to simplify the subsequent equations.  For completeness,
we also give the Gauss-Weingarten formulae in these variables:

\def\haf{{\scriptstyle{\scriptstyle 1\over\scriptstyle 2}}}

$$ \eqalign{
{\partial\over\partial x_1} \pmatrix \t_1\cr\t_2\cr\n\endpmatrix &=
\pmatrix
\haf g_{1\c1}&-\haf g_{1,2}e^{g_1-g_2}&e^{g_1}\kappa_1\cr
\haf g_{1,2}&\haf g_{2,1}&0\cr -\kappa_1&0&0
\endpmatrix
\pmatrix \t_1\cr\t_2\cr\n\endpmatrix \cr
\noalign{\vskip 12pt}
{\partial\over\partial x_2} \pmatrix \t_1\cr\t_2\cr\n\endpmatrix &=
\pmatrix
\haf g_{1,2}&\haf g_{2,1}&0\cr
-\haf g_{2,1}e^{g_2-g_1}&\haf g_{2,2}&e^{g_2}\kappa_2\cr0&-\kappa_2&0
\endpmatrix
\pmatrix \t_1\cr\t_2\cr\n\endpmatrix
}\eqno(2.2)$$

As the surface moves, the coordinate
lines will not remain principal coordinates unless
$\dot g_{12}=\dot h_{12}=0$. (This is where we begin to specify $V^\mu$
in order to maintain the desired coordinate system.)
This leads to two equations in $V^1_{\;\c2}$ and $V^2_{\;\c1}$ whose solution
is
$$
V^1_{\;\c2}=e^{-g_1}f,\qquad V^2_{\;\c1}=-e^{-g_2}f\eqno(2.3)$$
where
$$ f= {(U_{\c2}\kappa_2)_{\c1}-(U_{\c1}\kappa_1)_{\c2}\over
(\kappa_2-\kappa_1)^2}. $$

With these choices, (1.7) reduces to four coupled equations:
$$\eqalignno
{ \dot g_{i}&=-2\kappa_i U + V^j g_{i\c j} + 2 V^i_{\;\c i} &(2.4a) \cr
  \dot \kappa_i& = \kappa_i^2 U +V^j \kappa_{i\c j} + e^{-g_i/2}
(e^{-g_i/2}U_{\c i})_{\c i} + e^{-g_{\ii}}g_{i\c \ii}U_{\c \ii}/2&(2.4b)}$$
($(i,\ii)=(1,2)$ or $(2,1)$; no sum on $i$, $\ii$). These may be written as
$$
\eqalign{
&{Dg_{i}\over Dt}-2 V^i_{\;\c i}=-2\kappa_i U \cr
&{D\kappa_i\over Dt}=\kappa_i^2 U + U_{\c s_i s_i} + e^{-g_{\ii}}g_{i\c \ii}
U_{\c \ii}/2}
\eqno(2.5)$$
where $D/Dt$ is total derivative in the direction $-V^j\partial_j$
({\it i.e.}
the right hand side of (2.5) gives the evolution at ``unpushed'' particle
labels) and $s_i$ is arclength along $x_i$, ({\it i.e.}
$s_1=\int^{x_1}\exp({1\over2}g_1(\acute x_1,x_2))\,d\acute x_1$).

As a special case of these equations, we note that if $\kappa_2=\partial_2=0$
so that
the surface only depends on $x_1$, then $f = 0$ in (2.3), we can take
$V^2=0$, $V^1=W(x_1)\exp(-{1\over2}g_1)$ say,
and we recover the equations
for the motion of a plane curve [14]:
$$ \dot g=2W_{\c s}-2\kappa U\qquad \dot \kappa=\kappa^2 U +U_{\c
ss}+W\kappa_{\c s} $$

Now we return to the general case, to reduce (2.4) to two coupled equations
for the evolution of the two curvatures,  $\kappa_1$ and $\kappa_2$.  These
two equations, along with their compatibility condition, (1.3a), seem to be
the simplest intrinsic formulation possible for a general surface.

In (2.3), the equations for $V^\mu$,  there is still the freedom to specify two
arbitrary functions, each of one variable. We can use this freedom to
specify $g_i$, and hence the metric, completely.  Here is one way to
specify the metric: require
$$g_1(x_1,0,t)=g_2(0,x_2,t)=0\quad\forall\ t,\eqno(2.6)$$
so that $g_{11}=g_{22}=1$ along these two coordinate lines, which are
both parameterized by arclength (see Figure 1).
Let $v^\mu$ be two functions
of integration in (2.3), {\it i.e.,}
\def\`{\acute}
$$\eqalign{
V^1(x_1,x_2)&=+\int_0^{x_2}\exp(-g_1(x_1,\`x_2))f(x_1,\`x_2)\,d\`x_2 +
v^1(x_1)\cr
V^2(x_1,x_2)&=-\int_0^{x_1}\exp(-g_2(\`x_1,x_2))f(\`x_1,x_2)\,d\`x_1 +
v^2(x_2)}
$$
where we have suppressed the explicit $t$ dependence.
Then, from (2.4a) and choosing $v_i(0)=0$, enforcing (2.6) requires
$$ v_i' = (\kappa_i U - \haf g_{i\c \ii} V^{\ii})\big|_{x_{\ii}=0} $$
Now (2.1) can be integrated to define $g_i$ completely:
$$ g_1=\int_0^{x_2}{2\kappa_{1\c2}\over \kappa_2-\kappa_1}\,d\`x_2,\qquad
   g_2=-\int_0^{x_1}{2\kappa_{2\c1}\over \kappa_2-\kappa_1} \,d\`x_1.$$
Substituting these into (2.4b) yields two coupled equations for the evolution
of the two curvatures, as desired.

In the $\kappa_2=\partial_2=0$ case, $v_1'=\kappa_1 U$ and in this gauge,
$x_1$ is arclength, so we get $W = \int \kappa_1 U\,ds$ as expected.

\head 3. Surfaces of constant negative curvature
\endhead

{}From the wealth of special surfaces and coordinate systems available,
we consider a particular choice for which the equations of motion take
a particularly simple form.  At every point on a surface whose Gaussian (or
total) curvature $K$
is negative, the curvature tensor has two null vectors ({\it i.e.} vectors
$v^\mu$ which annihilate $h_{\mu\nu}$, so $v^\mu v^\nu h_{\mu\nu}=0$; see
[19, p. 70]).
An {\it asymptotic line} is a smooth curve
on the surface whose tangent vector coincides with one of these null
vectors at each point along the curve.  If the Gaussian curvature is
constant (in space) and negative, then there are two sets of asymptotic
lines, and they form a
{\it Tchebyshev net, i.e.}, they can be parameterized by arclength everywhere
[19, p. 368].
With these as coordinate lines, the
metric and curvature tensors take the form
$$ g_{\mu\nu}=\pmatrix 1&\cos \omega\cr \cos \omega&1\endpmatrix,
\qquad h_{\mu\nu}=\pmatrix 0&h_{12}\cr h_{12}&0\endpmatrix,\eqno(3.1)$$
where $\omega$ measures the angle between the coordinate lines.
Furthermore, the Gauss-Codazzi equations
become:
$$\eqalignno{ h_{12}&=\sqrt{-K}\sin \omega,&(3.2a)\cr
 \qquad \omega_{\c12}&=-K \sin \omega,&(3.2b)}$$
{\it i.e.}, solutions of the sine-Gordon equation are in a local 1--1
correspondence
with
such surfaces. Unfortunately such coordinates may be only local [19, p. 373].
The Gauss-Weingarten equations (1.2) also take the
simple form
$$ \eqalignno{
T_{\c1}&=\pmatrix \omega_{\c1}\cot \omega&-\omega_{\c1}\csc
\omega&0\cr0&0&\sqrt{-K}\sin \omega\cr
\sqrt{-K}\cot \omega&-\sqrt{-K}\csc \omega&0\endpmatrix T&(3.3)\cr
&=\pmatrix -(\ln(\kappa_1-\kappa_2))_{\c1}&-(\ln(\kappa_1+\kappa_2))_{\c1}&0\cr
0&0&-\kappa_1\kappa_2/(\kappa_1-\kappa_2)\cr
\kappa_1+\kappa_2&-(\kappa_1-\kappa_2)&0\cr \endpmatrix T }$$
and similarly for $T_{\c2}$, where
$$ T=\pmatrix \t_1\cr\t_2\cr\n\endpmatrix$$
and $\kappa_1$, $\kappa_2$ are the principal curvatures $\sqrt{-K}(\cot
\omega\pm\csc \omega)$.

The first step is to find those normal velocities $U$ that keep the Gaussian
curvature constant in space (but not necessarily in time).  Substitute
(3.1) into (1.7), and require both that the form of (3.1) be preserved
({\it i.e.,} $\dot g_{11}=\dot g_{22}=\dot h_{11}=\dot h_{22}=0$)
and consistency (so that $\dot\omega$ from (1.7a) equals that
from (1.7b)).   One finds in this way that $U$ must satisfy a linear
inhomogeneous equation:
$$ U_{\c12}=-KU\cos \omega+(\sin \omega){d\over dt}\sqrt{-K}\eqno(3.4) $$
and that the tangential components of velocity must satisfy:
$$ V^i_{\;\c \ii}=-(KU\sin \omega +\omega_{\c \ii}((-1)^{\ii}U_{\c \ii}\cos
\omega+U_{\c
i})+U_{\c \ii\ii}\sin \omega)/2\sqrt{-K}\sin^2\omega\eqno(3.5)$$
where $i$, $\ii = (1, 2)$ or $(2, 1)$, and there is no sum on $\ii$.  There are
no
other constraints.  As mentioned earlier, only the normal component of
velocity actually moves the surface, so we have established the following
\medskip
\noindent{\bf Proposition: \it
Suppose that a surface whose Gaussian curvature $K$ is
constant in space and negative moves according to {\rm (1.5)}.  It will
maintain
(spatially) constant negative curvature if and only if $U$ satisfies
{\rm (3.4)}.}
\medskip

Note that if $\dot K = 0$, then (3.4) is just the linearization of the
sine-Gordon equation, (3.2b), so its translational symmetry immediately
gives two solutions of (3.4): $U=\omega_{\c1}$  and $U=\omega_{\c2}$.
With the velocity components
chosen  to satisfy (3.4) and (3.5), the evolution of $\omega$ is given by
$${D\omega\over Dt}+(V^1_{\;\c2}+V^2_{\;\c1})\sin\omega=2\sqrt{-K}U\eqno(3.6)$$
where $D/Dt$ was defined below (2.5).

As an example, take $\dot K=0$ and $U=\omega_{\c 1}$ (=$\omega_{\c s_1}$).
Eqs. (3.5) can be integrated to find
$$V^1=((\omega_{\c 1})^2+2\omega_{\c 11}\cot \omega)/4\sqrt{-K},\qquad
  V^2=-\omega_{\c 11}\csc \omega/2\sqrt{-K}$$
so
$$\dot \omega={1\over
4}\left(6\sqrt{-K}\omega_{\c1}+\left((\omega_{\c1})^3+2\omega_{\c111}\right)/
\sqrt{-K} \right)\eqno{(3.7)}$$
which is the modified Korteweg-de Vries equation
in the variable $\omega_{\c1}$. The occurrence of an equation from
the AKNS hierarchy is not surprising, for the following reason. Let
$\omega(x_1,0)$
be an arbitrary function. It can be integrated in $x_2$ under (3.2b) to sweep
out a surface of constant negative curvature, or (because (3.7) does not
depend explicitly on $x_2$) it can be integrated in $t$ under (3.7). These
two AKNS evolution equations commute---thus (3.7) obviously preserves
the constant curvature property. In a sense, one should regard both $x_2$ and
$t$ as time-like variables. A similar instance of time-like evolution in
spatial directions was reported in [2].
It is not known whether more complicated
evolutions ({\it e.g.} $U=\omega_{\c1}+\omega_{\c2}$) can give rise to
integrable equations.

Now that it is known that this surface motion can be expressed as the motion
of a space curve it is fruitful to explore further. The coordinate
lines are asymptotic and hence their principal unit normals lie in the surface
(this can be seen from (3.3), because $T_{1\c1}$ is proportional to the
principal unit normal of the $x_1$-curve).
Furthermore
for the particles in a coordinate line to trace out the coordinate grid
properly it is easy to see (see Figure 2) that
$$\r_{\c2}=\sin \omega \,\n_c+\cos \omega\,\bt_c\equiv U_c\n_c+W_c\bt_c$$
where the subscript $_c$ denotes properties of an $x_1$-coordinate curve,
$\kappa_c$ is the curvature of the coordinate curve,
and $\tau_c$ is its torsion.
We know that this ``motion'' ({\it i.e.,} motion in the $x_2$-direction, as the
curve sweeps out the initial surface)
maintains the arc-length parameterization, which
requires $W_{\c1}=\kappa_c U\Rightarrow\kappa_c=-\omega_{\c1}$.
Now take this curve motion and find $\kappa_{c\c2}$ using the space-curve
evolution equations of [14]: the result is
$$ \kappa_{c\c2}=-\tau^2\sin\omega.$$
Because $\omega$ satisfies the sine-Gordon equation, we get $\tau=\sqrt{-K}$,
{\it i.e.,}
Tchebyshev coordinate lines in a surface of constant negative
curvature have constant torsion [5]. The curve is evolving under the equation
of motion
$$ \r_{\c2}= -\sin\smallint\kappa_c\,ds\, \n_c+
\cos\smallint\kappa_c\,ds\,\bt_c\eqno(3.8)$$

We have in fact recovered the constant-torsion sine-Gordon curve
equation of Lamb [12]. However, due to his inverse method (specifying
$\dot\bt$, $\dot\n$, $\dot\b$ instead of $\dot\r$), he was unable to
determine $\dot\r$ for this evolution. In the forward direction it is
easy to check that we have in fact the same evolution.

({\smc note:} In the literature there
are two distinct curve motions which obey the sine-Gordon equation.
Let the curvature evolve by $\dot\kappa_c\equiv \Omega U$. Lamb's
evolution, given by (3.8), has
$\Omega U=0$, and the curve would not move at all if $\tau_c=0$.
Nakayama {\it et al.} [14] found a {\it planar} curve motion which obeys
sine-Gordon: it has $\Omega^2 U=0$.)

As pointed out above, we can also think of an $x_1$-coordinate line as moving
in the $t$-direction, governed by the equation of motion (see Figure 2)
$$\eqalign{
\dot\r&=U\b_c+V^1\bt_c+V^2(\cos \omega\,\bt_c+\sin \omega\,\n_c)\cr
&=-\kappa_c\b_c+\kappa_{c\c1}\n_c/2\sqrt{-K}+\kappa_c^2\bt_c/4\sqrt{-K}
\cr
}\eqno(3.10)
$$
The final cross-check is that a space curve with motion given by (3.10)
has (from the curve equations in Nakayama {\it et al.} [14]) intrinsic
evolution
$$\eqalign{
\dot\kappa_c&={3\over2}\sqrt{-K}\kappa_{c\c s}+(3\kappa_c^2\kappa_{c\c s}+
2\kappa_{c\c sss})/4\sqrt{-K}\cr
\dot\tau_c&=0\cr}$$
agreeing with (3.7) when $\kappa_c=-\omega_{\c1}$.
\bigskip
{\eightpoint\baselineskip=10pt

\noindent
{\it Acknowledgements: } The authors are grateful for helpful conversations
with
S. Chakravarty, M. Mineev, K. Nakayama, and M. Wadati,
and to Nakayama and Wadati for
sharing their manuscript prior to publication.  This work was supported in
part by NSF grant \#DMS-9096156.
}

\vskip 1in
\parindent=0pt
\font\cs=cmcsc10
{\cs Figure 1.} Definition of the metric.

An origin is fixed and two coordinate lines are chosen
to be parameterized by arclength for all time; the metric at other
points such as $p$ is found by integrating the Gauss-Codazzi equations
along a coordinate line.
\vskip 1in

{\cs Figure 2.} Tchebyshev net on a surface of constant negative curvature.

All vectors shown are unit vectors, the normal $\bold n_c$ to the
coordinate curve lies in the surface, and the surface normal $\bold n$
is at right angles to it.
\vskip 1in


\Refs

\ref\no 1
\by	M. J. Ablowitz, D. J. Kaup, A. C. Newell, and H. Segur
\paper	The inverse scattering transform---Fourier analysis for
	nonlinear problems
\jour	Stud. Appl. Math. \vol 53 \yr 1974 \pages 249--315
\endref

\ref\no 2
\by	M. J. Ablowitz, R. Beals, and K. Tenenblat
\paper	On the solution of the generalized wave and generalized sine-Gordon
	equations
\jour	Stud. Appl. Math. \vol 74 \yr 1986 \pages 177--203;
	Addenda and corrigenda, {\bf 77} (1987) 101
\endref

\ref\no 3
\by	R. Brower, D. Kessler, J. Koplik, and H. Levine
\paper	Geometrical models of interface evolution
\jour	Phys. Rev. A \vol 29 \yr 1984 \pages 1335--1342
\endref

\ref\no 4
\by	J. D. Buckmaster and  G. S. S. Ludford
\book	Lectures on Mathematical Combustion,
CBMS-NSF Series in Applied Math.\vol 43
\publ	SIAM \publaddr Philadelphia \yr 1983
\endref

\ref\no 5
\by	L. P. Eisenhart
\book	An Introduction to Differential Geometry
\publ	Princeton University Press
\publaddr	Princeton
\yr	1947
\endref

\ref\no 6
\by	M. Gage and R. S. Hamilton
\paper	The heat equation shrinking convex plane curves
\jour	J. Diff. Geom. \vol 23 \yr 1986 \pages 69--96
\endref

\ref\no 7
\by	R. E. Goldstein and D. M. Petrich
\paper	The Korteweg-de Vries hierarchy as dynamics of closed curves
	in the plane
\jour	Phys. Rev. Lett. \vol 67 \yr 1991 \page 3203--3206
\endref

\ref\no 8
\by	M. Grayson
\paper	The heat equations shrinks embedded plane curves to round points
\jour	J. Diff. Geom.\vol 26\yr 1987\pages 285--314
\endref

\ref\no 9
\by	Hasimoto
\paper	A soliton on a vortex filament
\jour	J. Fluid Mech. \vol 51 \yr 1972 \pages 477--485
\endref

\ref\no 10
\by   	G. Huisken
\paper	Flow by mean curvature of convex surfaces into spheres
\jour 	J. Diff. Geom. \vol 20 \yr 1984 \pages 237--266
\endref

\ref\no 11
\by	D. A. Kessler, J. Koplik, and H. Levine
\paper	Pattern selection in fingered growth phenomena
\jour	Adv. Phys. \vol 37 \yr 1988\pages 255--339
\endref

\ref\no 12
\by	G. L. Lamb, Jr.
\paper	Solitons on moving space curves
\jour	J. Math. Phys. \vol 18 \yr 1977 \pages 1654--1661
\endref

\ref\no 13
\by	J. S. Langer
\paper	Lectures in the theory of pattern formation
\inbook	in Chance and Matter
\pages	629--712
\eds	J. Soutelie, J. Vannimenus, and R. Stora
\publ	North-Holland \publaddr New York \yr 1987
\endref

\ref\no 14
\by	K. Nakayama, H. Segur, and M. Wadati
\paper	Integrability and the motion of curves
\jour	Phys. Rev. Lett. \vol 69 \yr 1992 \pages 2603--2606
\endref

\ref\no 15
\by	K. Nakayama and M. Wadati
\paper	The motion of surfaces
\jour	preprint \yr 1992
\endref

\ref\no 16
\by	P. Pelc\'e
\book	Dynamics of Curved Fronts
\publ	Academic Press \publaddr New York \yr 1988
\endref

\ref\no 17
\by	J. A. Sethian
\paper	Numerical algorithms for propagating interfaces: Hamilton-Jacobi
equations and conservation laws
\jour	J. Diff. Geom.\vol 31\yr 1990\pages 131--161
\endref

\ref\no 18
\by	J. Smoller
\book	Schock Waves and Reaction-Diffusion Equations
\publ	Springer
\publaddr	New York
\yr	1980
\endref

\ref\no 19
\by     M. Spivak
\book   A Comprehensive Introduction to Differential Geometry, Volume III
\publ   Publish or Perish, Inc.
\publaddr       Boston
\yr     1975
\endref

\ref\no 20
\by	J. J. Stoker
\book	Water Waves
\publ	Wiley-Interscience \publaddr New York \yr 1957
\endref

\endRefs
\enddocument